\documentstyle[preprint,aps]{revtex}

\begin{document}
\draft
\title{K-quantum Nonlinear Jaynes-Cummings Model in Two Trapped Ions \thanks{%
This work was supported by the National Natural Science Foundation of China
(19734006 and 69873015) and Chinese Academy of Science}}
\author{Hao-Sheng Zeng$^{1,2}$}
\address{$^1$Laboratory of Magnetic Resonance and Atomic and Molecular\\
Physics, Wuhan Institute of Physics and Mathematics, Chinese Academy of\\
Science, Wuhan 430071, People's Republic of China \\
$^2$Department of Physics, Hunan Normal University, Hunan\\
410081, People's Republic of China}
\maketitle

\begin{abstract}
A k-quantum nonlinear Jaynes-Cummings model for two trapped ions interacting
with laser beams resonant to k-th red side-band of center-of-mass mode, far
from Lamb-Dicke regime, has been obtained. The exact analytic solution
showed the existence of quantum collapses and revivals of the occupation of
two atoms.
\end{abstract}


\vskip 1cm

\narrowtext

In quantum optics, Jaynes-Cummings model(JCM) [1] has been regarded as a
fundamental research model which can be exactly solved and predicts some
interesting nonclassical effects both with respect to the atomic states and
the boson mode, such as atomic revivals[2] and squeezing[3]. Experimental
realization of JCM with dissipation has been demonstrated both in microwave
and in optical domain. Observation of quantum revivals and nonclassical
photon statistics with Rydberg atoms and microwave cavities has been
reported [4], and in the optical regime vacuum Rabi splitting has been
observed [5]. This experimental progress stimulated a manifold of further
theoretical studies, for example of the quantum correlation between two
interacting subsystems [6]. Moreover, related models have been rendered
successively, such as nonlinearly coupled [7] and multiquantum JCMs [8],
Raman-type [9], and three-level models [10].

Trapped ions interacting with laser beam is another fundamental system to
exhibit, in appropriate limits, JCM dynamics. The quantized center-of-mass
motion of the ion in the trap potential plays the role of the boson mode,
which is coupled via the laser to the internal degrees of freedom. An
analogy between an undamped trapped ion and the JCM has been noted by
Blockley, Wall, and Risken [11], and by J. I. Cirac, et al., for damped
trapped ion [12]. A study for nonlinear multiquantum JCM dynamics of a
trapped ion is presented by W. Vogel and R. L. de Matos Filho [13]. A
significant feature of trapped-ion JCM is that the effective coupling
constant, which depends on Rabi frequency, could be readily adjusted in
experiments to satisfy the strong coupling condition.

In recent experiments of trapped ions, entangled states up to four ions have
been realized, and some applications, such as the proof of Bell's inequality
and decoherence-free quantum memory, have been confirmed[14]. These
improvements offered some advantageous conditions for studying properties of
multi-ion JCM.

In this paper we will demonstrate that a system comprising two trapped ions
interacting with a laser beam resonant to k-th red side-band of
center-of-mass mode, far from Lamb-Dicke regime, shows a nonlinear JCM
dynamics. Collapses and revivals can be observed in this system. These
properties of internal states for two trapped ions are useful for quantum
computation. Because quantum manipulation involves at lest two-bit
operations, and only we know the properties of internal state of ions very
clearly, can we manipulate them well and truly. In addition, the model for
this system may be realized in present experimental conditions[14], which
will result in directly the tests of the predictions from this paper.

Consider two ions trapped in a linear trap which are strongly bounded in the 
$y$ and $z$ directions but weakly bounded in a harmonic potential in $x$
direction.. Assuming that the two ions are illuminated simultaneously with a
dispersive beam resonant with k-th red side-band of center-of-mass mode, so
that the effects from the relative motional mode can be neglected because of
the very large off-resonant reason[14,15]. The Hamiltonian for this system is

\begin{equation}
H=H_0+H_{int},
\end{equation}

\begin{equation}
H_0=\nu (a^{+}a+1/2)+\omega \sum_{i=1}^2\sigma _{iz}/2,
\end{equation}

\begin{equation}
H_{int}=\sum_{i=1}^{2}\frac{\Omega }{2}\left\{ \sigma _{+i}e^{i\eta
(a+a^{+})}e^{-i(\omega -k\nu )t}+H.C.\right\}
\end{equation}

where $\nu $ and $a^{+}$ ($a$) are the frequency and ladder operators of the
center-of-mass mode. $\omega $ is the energy difference between the ground
state $\left| 0\right\rangle $ and the long-lived metastable excited state $%
\left| 1\right\rangle $ of the ion. For simplicity we further assume that
the Rabi frequency and Lamb-Dicke parameter for the two ions are same.
Transforming the Hamiltonian (3) to the interaction picture with respect to $%
H_{0}$ and making use of rotating wave approximation, we have,

\begin{equation}
H_{int}=\Omega J_{+}f(a^{+}a)a^{k}+H.C.
\end{equation}
with

\begin{equation}
f(a^{+}a)=e^{-\eta ^{2}/2}\sum_{n=0}^{\infty }\frac{(i\eta )^{2n+k}}{n!(n+k)!%
}(a^{+})^{n}a^{n}
\end{equation}
where $J_{\pm }=J_{x}{\pm }iJ_{y}$ with $J_{\alpha }=\frac{1}{2}(\sigma
_{1\alpha }+\sigma _{2\alpha })$ ($\alpha =x,~y,~z$) being the three
components of total spin operator of the two ions.

E.q.(4) is namely the so-called k-quantum nonlinear JCM Hamiltonian. When
any one of the two ions is excited, the quantum number of $J_{z}$ component
of the total spin operator increases by one, and in the meantime, the
quantum number of center-of-mass mode decreases by $k$.

Following the method of [16], we can get the exact solution of Hamiltonian
(4) after a tedious deduction,

\begin{equation}
U_{11}=\frac{1}{A^{2}+B^{2}}\left[ A^{2}\cos (\sqrt{A^{2}+B^{2}}%
t)+B^{2}\right]
\end{equation}
\begin{equation}
U_{00}=\cos (\sqrt{A^{2}+B^{2}}t)
\end{equation}
\begin{equation}
U_{-1-1}=\frac{1}{A^{2}+B^{2}}\left[ B^{2}\cos (\sqrt{A^{2}+B^{2}}%
t)+A^{2}\right]
\end{equation}
\begin{equation}
U_{10}=-U_{01}^{*}=\frac{-i^{k+1}A}{\sqrt{A^{2}+B^{2}}}\sin (\sqrt{%
A^{2}+B^{2}}t)
\end{equation}
\begin{equation}
U_{1-1}=U_{-11}=\left( -1\right) ^{k+1}\frac{AB}{A^{2}+B^{2}}\left[ 1-\cos (%
\sqrt{A^{2}+B^{2}}t)\right]
\end{equation}
\begin{equation}
U_{0-1}=-U_{-10}^{*}=\frac{-i^{k+1}B}{\sqrt{A^{2}+B^{2}}}\sin (\sqrt{%
A^{2}+B^{2}}t)
\end{equation}
where $A=\sqrt{2}\Omega e^{-\eta ^{2}/2}\eta ^{k}\sqrt{\frac{\left(
n-2k\right) !}{\left( n-k\right) !}}L_{\left( n-2k\right) }^{k}\left( \eta
^{2}\right) $ and $B=\sqrt{2}\Omega e^{-\eta ^{2}/2}\eta ^{k}\sqrt{\frac{%
\left( n-k\right) !}{n!}}L_{\left( n-k\right) }^{k}\left( \eta ^{2}\right) $
with $L_{n}^{k}\left( x\right) =\sum_{m=0}^{n}\left( -1\right) ^{m}\left( 
\begin{array}{l}
n+k \\ 
n-m
\end{array}
\right) \frac{x^{m}}{m!}$.

In order to study the phenomena of collapses and revivals of the two ions,
imitating traditional process, we assume that initially the internal states
of the two ions are both in the ground states and the external motional
state of center-of-mass mode in a coherent state $\mid \alpha \rangle $.
Thus the whole initial state of the system is,

\begin{equation}
\rho(0)=\sum_{m^\prime, m^{\prime \prime }=0}^\infty q(m^\prime) q^*
(m^{\prime \prime }) \mid -1~ m^\prime \rangle \langle -1~ m^{\prime \prime}
\mid
\end{equation}
with $q(m)=exp(-\frac{1}{2} \mid \alpha \mid^2 ) \frac{\alpha ^m}{\sqrt m!} $
are the coefficients for the coherent state expanding in the number state
basis.

The density operator of the internal state of this system can be obtained by
tracing over the vibrational motion of center-of-mass mode. With above
equations, we can get the time evolution of diagonal elements of the
density-matrix for the two-ion system,

\begin{equation}
\rho _{11}(t)=\sum_{n=2}^{\infty }\frac{A^{2}B^{2}}{(A^{2}+B^{2})^{2}}\left[
1-\cos (\sqrt{A^{2}+B^{2}}t)\right] ^{2}p\left( n\right)
\end{equation}

\begin{equation}
\rho _{00}(t)=\sum_{n=2}^{\infty }\frac{B^{2}}{A^{2}+B^{2}}\left[ \sin (%
\sqrt{A^{2}+B^{2}}t)\right] ^{2}p(n)+\left[ \sin B\left( 1\right) t\right]
^{2}p\left( 1\right)
\end{equation}

\begin{equation}
\rho _{-1-1}(t)=\sum_{n=2}^{\infty }\frac{1}{\left[ A^{2}+B^{2}\right] }%
\left[ A^{2}+B^{2}\cos (\sqrt{A^{2}+B^{2}}t)\right] ^{2}p(n)+\left[ \cos
B\left( 1\right) t\right] ^{2}p\left( 1\right) +p\left( 0\right)
\end{equation}
with $B(1)=\sqrt{2}\Omega e^{-\eta ^{2}/2}\eta $ and 
\begin{equation}
p(n)=e^{-\mid \alpha \mid ^{2}}\frac{\mid \alpha \mid ^{2n}}{n!}
\end{equation}
is the probability distribution of phonon number for coherent state of
center-of-mass mode.

According to E.q.(13)-(15), we plotted the exact time-evolution of diagonal
elements of density-matrix as in Fig.1, which describes the time-evolution
of occupation probabilities of electronic energy-levels. we assumed that
initially the internal states of the two ions are both in ground, and the
vibration motion of center-of-mass mode is in a coherent state with average
phonon number $\mid \alpha \mid ^{2}=10$. The typical parameters are $\eta
=0.1$ and $\Omega =2\pi \times 500$ kHz [14]. From these curves, we can see
clearly the collapses and revivals for the electronic state occupation of
the two trapped ions. It is worth while to point out that, in this system,
collapses and revivals increase with the increase of average phonon number,
but decrease with the increase of Lamb-Dicke. In order to observe collapses
and revivals in the case of large Lamb-Dicke parameters, we must enhance
average phonon number. As a comparison, we plotted the similar figure with
the same Rabi frequency and parameters $\eta =0.2$, $\mid \alpha \mid
^{2}=50 $ and $\eta =0.4$, $\mid \alpha \mid ^{2}=80$ in Fig. 2. Obviously
the collapses and revivals in Fig.2 (b) is very obscure, because of the very
large Lamb-Dicke parameter.

Collapses and revivals in two-quantum resonance are very similar to the case
of one-quantum resonance. The occupation distribution of internal
energy-level for different Lamb-Dicke parameters and different average
phonon number for this case are shown in Fig.3(the Rabi frequency is same as
above.).

In conclusion, we showed that the behavior of two trapped ions, interacting
with a laser beam resonant to k-th red sand-band of center-of-mass mode, can
be described by a nonlinear JCM mode under appropriate limits. An exact
analytic solution for this type of JCM was presented and the results
indicated obvious collapses and revivals. We expect the similar behaviors
for more than two ions, and also expect new features of the nonclassical
statistics of the vibrational mode and of other effects known from the
standards JCMs.

Fig.1. The occupations $\rho _{11}(t),~\rho _{00}(t),~\rho _{-1-1}(t)$ of
electronic state of the two ions for one-quantum resonance are given as
functions of time with parameters: $\eta =0.1$, $\Omega =2\pi \times 500$
kHz. Initially the motional state of center-of-mass mode be in a coherent
state with average phonon number $\mid \alpha \mid ^{2}=10$ and the two ions
be in ground state.

Fig.2. The time-evolution of the occupations $\rho _{11}(t)$ of excited
electronic state of the two ions for one-quantum resonance. (a) $\eta =0.2$, 
$\mid \alpha \mid ^{2}=50$; (b) $\eta =0.4$ $\mid \alpha \mid
^{2}=\allowbreak 80$.

Fig.3. The occupations $\rho _{11}(t)$ of excited electronic state of the
two ions for two-quantum resonance are given as functions of time with
parameters: (a) $\eta =0.1$, $\mid \alpha \mid ^{2}=20$; (b) $\eta =0.2$, $%
\mid \alpha \mid ^{2}=50$; (c) $\eta =0.4$, $\mid \alpha \mid ^{2}=80$.

\end{document}